# N-body Information Theory (NbIT) Analysis of Rigid-Body Dynamics in Intracellular Loop 2 of the 5-HT$_{2A}$ Receptor

Michael LeVine[1], Jose Manuel Perez-Aguilar[1], Harel Weinstein[1,2]

[1] Department of Physiology and Biophysics, Weill Medical College of Cornell University, New York, New York, USA.
`{mil2037, jop2032}@med.cornell.edu`
[2] The HRH Prince Alwaleed Bin Talal Bin Abdulaziz Alsaud Institute for Computational Biomedicine, Weill Medical College of Cornell University, New York, New York, USA.
`haw2002@med.cornell.edu`

**Abstract.** Rigid-body motions of protein secondary structure are often implicated in mechanisms of protein function. In GPCRs, evidence suggests that intracellular loop 2 (IL2) contains a segment characterized as a helix when the activated receptor triggers downstream signaling. However, neither experimental nor computational methods are readily available to assess quantitatively the degree of collective motions in such secondary structure motifs of proteins. Here we describe a new element of our N-body Information Theory (NbIT) framework to address this problem. To this end we introduce *total intercorrelation*, a measure in information theory that can be used to describe n-body correlated dynamics between multivariate distributions, such as 3-dimensional atomic fluctuations in simulations of proteins. We also define two additional measures, the *rigid-body fraction* and *correlation order*, which can be determined from the decomposition of the configurational entropy. Using these measures, we analyze the dynamics of IL2 in microsecond Molecular Dynamics simulations of the 5-HT$_{2A}$ receptor to demonstrate the powerful features of the new analysis techniques in studying the collective motion dynamics of secondary structure motifs. The analysis reveals an intriguing difference in the extent of correlated motions in the helical segment of IL2 in the presence and absence of bound 5-HT, the endogenous agonist that activates the receptor and triggers downstream signaling, suggesting that IL2 rigid-body motions can display distinct behaviors that may discriminate functional mechanism of GPCRs.

## 1 Introduction

There is sustained and longstanding interest in the involvement of specific secondary and tertiary structure elements in the function of signaling proteins at the cell surface. In particular, the coordinated movements of these structural components in molecular machines, such as the G-protein-coupled receptors (GPCRs), have been proposed as key mechanistic elements, and their properties and dynamics have been examined experimentally and computationally for a long time. Indeed, the hypothet-ical mechanisms for transitions between functional states of molecular machines that have been determined structurally, *e.g.*, with x-ray crystallography, are often considered in terms of rigid-body movements of elements with determined secondary structure content. But the experimental validation, at atomic resolution, of the extent to which such rigid-body motions are realistic and actually involved in specific mechanisms, remains impractical because the molecular machines contain a very large number of such higher-order structural elements. Thus, even within a single crystallographic state, domains are often considered to be "rigid" if the experimental B-factor is low, and "disordered" if the B-factor is high, although it would be necessary to validate such statements by examining the many-body correlated motions that define a true rigid body. In fact, a more quantitative evaluation of such many-body correlations may reveal large divergence from "rigid-body" behavior in the secondary structure elements; conversely, loops and coil regions, often viewed as "disordered", may incorporate hidden collective motions that could be essential for their role in the functional properties of the entire protein.

Recently it has been proposed that transitions between structured states of the intracellular loop 2 (IL2) in GPCRs (*i.e.*, including a helix capable of rigid-body motions), and unstructured states of the loop, may be involved in signaling mechanisms of the receptor. In crystal structures of the GPCRs in their inactive states, *e.g.*, the $\beta_2$ adrenergic receptor ($\beta$2AR) bound to the inverse agonist carazolol (2RH1.pdb) [1] and inactivated rhodopsin (1F88.pdb) [2], intracellular loop 2 (IL2) is unstructured; but IL2 is structured in the crystal structures of the active states of the same receptors, *e.g.*, the $\beta$2AR bound to the agonist BI-167107 (3SN6.pdb) [3] and the structure of metarhodopsin II (3PQR.pdb) [4]. Moreover, mutations in IL2 that affect receptor function have been documented [5–7], and many of these were predicted to disrupt the conformation and/or helicity of IL2 [8, 9]. The inferences from these studies were recently bolstered by the determination of crystal structures of two serotonin (5-HT) receptors, 5-HT$_{1B}$R [10] and 5-HT$_{2B}$R [11], bound to the agonist ergotamine. Ergotamine binding to 5-HT$_{1B}$R activates the receptor for signaling through G-protein- or $\beta$-arrestin-dependent pathways. In contrast, while the 5-HT$_{2B}$R bound to ergotamine can still signal through both pathways, the coupling to G-protein-dependent pathways is greatly reduced [11]. Interestingly, the structure of ergotamine-bound 5-HT$_{1B}$R shows a helical motif in IL2 but is unstructured in the ergotamine-bound 5-HT$_{2B}$R (see Fig. 1). Thus, if the presence of a helix in IL2 is preferred for the activation of G-protein signaling, the lack of such a helix in the IL2 of 5-HT$_{2B}$R bound to ergotamine may explain why it displays impaired G-protein-dependent signaling. While these results may indicate that the unstructured conformation of IL2 inhibits the activation of some downstream signaling, they do not speak to the role of agonist in modulating the conformation of IL2 and are limited to phenomenological helical/non-helical binary classification. A quantitative measure of collective motions is thus needed.

To generate a quantitative measure of the rigid-body properties in protein segments such as the IL2, we have developed methods based on the configurational entropies obtained from unbiased all-atom Molecular Dynamics (MD) simulations of the molecular systems. The new methods are illustrated here for a GPCR, the serotonin 2A receptor, 5-HT$_{2A}$R. Developed within the previously introduced framework based on information theory, N-body Information Theory (NbIT) analysis [12], the present work accomplishes two goals. First, we introduce a new measure of information that corrects previous problems encountered when characterizing the correlation between multivariate distributions using *mutual information,* and show that our new measure is better able to capture the properties of rigid-body systems. Then we take advantage of the decomposition of the system's configurational entropy as an N-order expansion of n-body information terms to calculate rigid-body parameters, and use these parameters to characterize IL2 in the apo and 5-HT bound state of the $5\text{-HT}_{2A}R$. Our discussion underscores how such detailed quantitative findings concerning the dynamics of specific structural elements, which are achieved as described here by using the new information theory-based analysis of MD trajectories, provide novel insights into modulation of local conformations (of the IL2 loop in this case) by ligands, and inform on the potential role of such modulation in the protein's function.

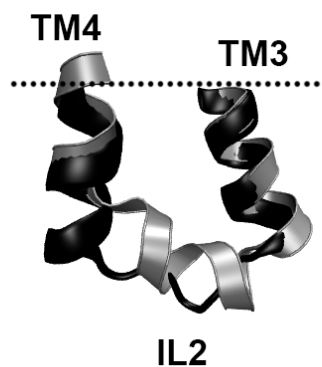

**Fig. 1.** IL2 contains a helical segment in $5\text{-HT}_{1B}R$, but an unstructured segment in $5\text{-HT}_{2B}R$. IL2 and the adjacent TMs, truncated at the dotted line, are shown. $5\text{-HT}_{1B}R$ is in silver ribbon representation and $5\text{-HT}_{2B}R$ is in black cartoon representation.

## 2  Theory

### 2.1  Total Intercorrelation

The trajectories collected from molecular dynamics (MD) simulations provide detailed atomistic insights into the dynamics of biological macromolecules and their equilibrium ensemble of conformations, which are nearly unattainable using current experimental methods. In particular, as a rigid-body system of particles would have low conformational entropy relative to a set of independent particles, the rigid-body properties can be calculated from the equilibrium ensemble of conformations. This becomes important for the analysis of the dynamics of specific structural elements of the biological macromolecules, which are considered to be folded into elements of secondary structure (*e.g.*, helices) that would move as rigid bodies. If the system is behaving like a rigid body, all particles in the system will be correlated and share a large amount of *mutual information,* which can be quantified with methods of Infor-mation Theory. Indeed, many have used *mutual information* to quantify pair-wise correlation in molecular systems [13–17]. *Mutual information* between two variables is defined as:

$$I_2(X_1, X_2) = H(X_1) + H(X_2) - H(X_1, X_2) \quad (1)$$

Here, $H(X_1)$ and $H(X_2)$ are the marginal entropy of variables $X_1$ and $X_2$ respectively and $H(X_1, X_2)$ is the joint entropy. The n-body information, the information shared by n variables, can be calculated recursively:

$$I_n(X_1, \ldots, X_n) = I_{n-1}(X_1, \ldots, X_{n-1}) - I_{n-1}(X_1, \ldots, X_{n-1} | X_n) \quad (2)$$

If an N-body system behaves as a rigid body, any information regarding one body is shared with all other bodies, and $I_N$ will be high.

However, the use of *mutual information* to describe correlations within atomic fluctuations in 3-dimensional space makes the erroneous assumption that *mutual information* measures the same correlation or dependency as the Pearson's correlation coefficient. In fact, in the multivariate case, *mutual information* describes a very different type of dependence due to the effects of correlations between internal dimensions of a multivariate distribution. In order to appreciate the effect of correlation between internal dimensions on *mutual information*, it is important to discuss first *multivariate entropy*. For any distribution of multivariate random vectors, $X = \{X_1, \ldots, X_d\}$, the marginal entropy of that distribution is the joint entropy of the d dimensions, $H(X) = H(X_1, \ldots, X_d)$. If the internal dimensions are not independent, the joint entropy is less than the sum of the marginal entropy of the internal dimensions, and that difference is the *total correlation*, TC:

$$TC(X) = \sum_{i=1}^{d} H(X_i) - H(X) \quad (3)$$

If a multivariate distribution contains *total correlation*, the joint entropy is decreased and thus the maximum information that can be shared with other distributions is decreased proportionately. For example, in a system of two atoms, A and B, where each coordinate has marginal entropy of H, if the coordinates are completely dependent, the joint entropy of each atom is also H and the maximum 2-body *mutual information*, $I_2(A, B)$, is H. However, if the coordinates are completely independent, the joint entropy of each atom is 3H and the maximum $I_2(A, B)$ is 3H. In this case:

$$\max(I_2(A, B)) = H * \min(d_A, d_B) \quad (4)$$

The above property of *mutual information* penalizes multi-body correlations between dimensions of different atoms. If the x dimension of atom A, $A_x$, is correlated with both the y and z dimensions of atom B, $B_y$ and $B_z$, there is 3-body information, $I_3(A_x, B_y, B_z)$. As a result, $I_2(B_y, B_z)$ increases and $H_B$ decreases, leading to a decrease in $I_2(A, B)$. As the 3-body information is part of the correlation we desire to quantify for the results of MD simulations of molecular systems at atomic resolution, it is clear that *mutual information* is not suitable for quantifying the correlation between multivariate distributions.

Here, we formulate a new correlation measure, *total intercorrelation*, which is a generalization of the *mutual information* of two 1-dimensional distributions and provides a better description of *multivariate correlation*. First we define this measure and then demonstrate its properties. We begin with the *total correlation* between two distributions A and B of dimension $d_A$ and $d_B$:

$$TC(A, B) = \sum_{i=1}^{d_A} H(A_i) + \sum_{i=1}^{d_B} H(B_i) - H(A, B) \quad (5)$$

As a measure, the *total correlation* includes internal correlations within each atom that are not part of higher n-body correlations with the other atom. In order to remove these correlations, we calculate the total correlation for each distribution, conditional on the other, and subtract it from the total correlation between all dimensions of both distributions to yield the *total intercorrelation*:

$$TC_{INTER}(A, B) = TC(A, B) - TC(A|B) - TC(B|A) \quad (6)$$

The *total intercorrelation* describes the total amount of information shared between two multivariate distributions through any n-body correlation that contains at least one dimension from both distributions, and equals $I_2(A, B)$ in the univariate case. *Total intercorrelation* is distinctly different from the ($d_A + d_B$)-body *mutual information* as it counts n-body information between dimensions of A and B that is not shared by all dimensions of both A and B.

For illustration, we discuss a system of two atoms where the marginal entropy of each dimension of each atom has been standardized to H. If all dimensions are perfectly coupled (i.e. knowing the value of one variable reveals the value of all variables):

$$max(TC_{INTER}(A, B)) = max(TC(A, B)) = H * (d_A + d_B - 1) \quad (7)$$

Because the maximum $TC_{INTER}$ scales with the $d_A + d_B - 1$, we relate the $TC_{INTER}$ between two distributions of any dimensions to the $TC_{INTER}$ between two 1-dimensional distributions by dividing by $d_A + d_B - 1$. Importantly, total correlation is a Kullback-Leibler divergence [18] and thus invariant to scaling of the distribution, so if each variable follows the same distribution, and that distribution can be standardized (such the multivariate normal distribution), the standardization condition is conveniently met without any transformation of the data.

*Total intercorrelation* can be generalized to a set of N bodies of arbitrary dimensions, $\{X_1, ..., X_N\}$, where $X_i = \{X_{i,1}, ..., X_{i,d_{X_i}}\}$. For an arbitrary N-body problem, the *total intercorrelation* can be calculated as:

$$TC_{INTER}(X_1, ..., X_N) = TC_{INTER}(X_1, ..., X_{N-1}) - TC_{INTER}(X_1, ..., X_{N-1}|X_N) \quad (8)$$
$$+ TMI(X_N, \{X_1, ..., X_{N-1}\})$$

TMI is the *total marginal information*, the sum of the information shared between each dimension of $X_N$ with all distributions in the set $\{X_1, ..., X_{N-1}\}$. TMI can also be calculated recursively, where:

$$TMI(X_N, \{X_1, \ldots, X_{N-1}\}) = TMI(X_N, \{X_1, \ldots, X_{N-2}\}) \qquad (9)$$
$$-TMI(X_N, \{X_1, \ldots, X_{N-2}\}|X_{N-1})$$

and:

$$TMI(X_2, X_1) = \sum_{i=1}^{d_{X_2}} H(X_{2,i}) - \sum_{i=1}^{d_{X_2}} H(X_{2,i}|X_1) \qquad (10)$$

Because the *total intercorrelation* between two 1-dimensional distributions is the *mutual information*, the N-body *total intercorrelation* between N 1-dimensional distributions is the N-body *mutual information* when N is greater than 2.

Using the N-body *total intercorrelation*, the total amount of rigid-body behavior can be quantified. We calculate a *generalized correlation coefficient*, r, [13]:

$$r = \sqrt{1 - e^{\frac{-2*I}{s(d)}}} \qquad (11)$$

I is an arbitrary information measure, and s(d) is the appropriate scaling coefficient as a function of the dimension which allows us to compare the information measure to the mutual information between two normal distributions with unit variance (for example, $\sum_{i=1}^{N} d_{X_i} - 1$ for $TC_{INTER}$ and $\min(d_{X_1}, \ldots, d_{X_N})$ for *mutual information*). We will use r as a function of *mutual information* ($r_{mutual}$) and *total intercorrelation* ($r_{inter}$) to investigate the overall amount of collective behavior.

We next introduce a method for investigating other rigid-body characteristics that provide additional detail using the N-body expansion of the configurational entropy.

### 2.2 Using the N-body Expansion of the Configurational Entropy to Define Rigid-Body Dynamics

The entropy of a system can be written as an expansion of the n-body information in n:

$$H = \sum I_1 - \sum I_2 + \sum I_3 - \sum I_4 \ldots \sum I_N \qquad (12)$$

Here each term $\sum I_n$ is the sum of all possible n-body *mutual information* terms, where $I_1$ corresponds to the marginal entropy. If the system is completely disordered, all terms n > 1 are approximately zero, and the entropy can be approximated by terminating the expansion at n = 1. If the system is completely ordered, corresponding to a rigid body, $\sum I_n = I_N * \binom{N}{n}$ and the entropy will oscillate according to the binomial coefficient such that it can only be accurately calculated at n = N.

Based on this, one can define two rigid-body parameters. One is the *rigid-body fraction* (RBF) that describes the average fraction of 2-body *mutual information* that is part of the rigid-body N-body motion.

$$RBF = \frac{I_N}{\langle I_2 \rangle} \qquad (13)$$

If all correlations are due to rigid-body motions, $\langle I_2 \rangle = I_N$ and RBF = 1. If there are no rigid-body motions, $I_N = 0$ and thus RBF = 0. For a completely disordered system, $\langle I_2 \rangle = I_N = 0$, and RBF is undefined.

In addition to quantifying the absolute rigid-body behavior, the existence of other n-body correlations that are greater than 2 but less than N can be revealed by how quickly the expansion of the entropy converges. We parameterize a function to describe the average n-body information term as a function of n from 2 to N, with the following exponential form:

$$I_n = Ae^{\frac{-(n-2)}{CO}} + B \qquad (14)$$

For a model system consisting of a finite 1-dimensional lattice of one-dimensional normal distributions with unit variance and uniform covariance between neighbors, we find that an approximately exponential decay is expected for a range of covariances. Additionally, adding heterogeneity by modifying some of the covariances did not change the decay (see Fig. 2).

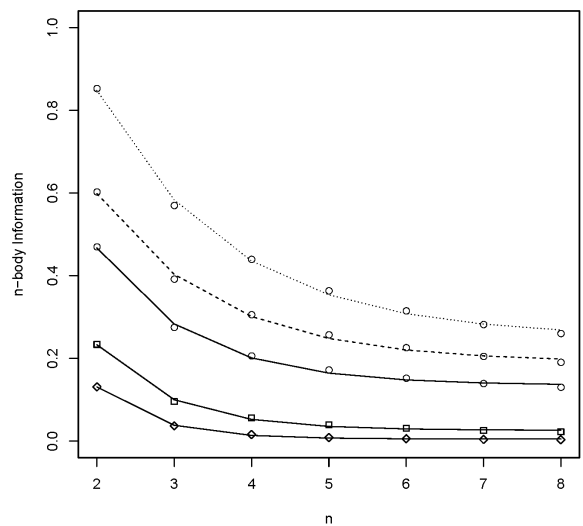

**Fig. 2.** Approximately exponential decay of n-body information in model 1-dimensional lattices of coupled 1-dimensional normal distributions. Covariance between neighbors are 0.7 (diamond), 0.8 (square), and 0.9 (circle, solid line). The dashed and dotted lines correspond to systems where the fourth and fifth distributions have greater covariance with their neighbors (0.95 and 0.99, respectively). All lines are parameterized as exponential decays using Eq. 14.

By parameterizing the exponential function, we calculate the *correlation order*, CO, which describes the rate of decay of n-body correlations in the system. A CO of 1 would indicate that the average nat of n-body information is contributed by an n + 1 body correlation. In the model system described above, if the correlations between neighboring distributions are low, the exponential will decay quickly and have a low CO, and if the correlations between neighbors are high, the exponential decays slowly due to the emergence of higher correlations, and have a high CO.

While the behavior of 1-dimensional lattices of 1-dimensional distributions is relatively simple, the dynamics of a protein are not expected to be as simple, as correlation occurs across 3 dimensions and are highly heterogeneous. CO is clearly a function of N, $\langle I_2 \rangle$ and $I_N$, but this will not be the case with higher dimension distributions, where the rigid-body motion corresponding to $I_N$ could occur in one dimension while other correlated motions occur in other dimensions. However, we find that for the secondary structure of proteins, which is investigated here, under the multivariate normal approximation, an approximate exponential decay of $I_n$ with n is still observed. Also, in addition to using the decay in n-body *mutual information* as a function of n to characterize rigid-body motions, it is also possible to use the decay in n-body *total intercorrelation* to reveal rigid-body motions, with the advantage of correcting for multi-body correlations between the Cartesian coordinates of atoms. We find that for the secondary structure studied here, n-*body total intercorrelation* also decays approximately exponentially.

## 3 Methods

### 3.1 Molecular Dynamics Simulations

Details pertaining to the construction and simulation of the molecular systems analyzed here can be found in [19]. Briefly, models of the apo and 5-HT bound 5-HT$_{2A}$R in an explicit membrane solvated with water and 0.15 M NaCl were studied with unbiased MD simulations. A 1 μs trajectory was collected with a 2 femtosecond time step at constant temperature (310 K) and pressure (1 atm) using the NAMD package [20] with the all-atom CHARMM27 force field with CMAP corrections [21].

### 3.2 Calculation of Entropy

Configurational entropy was calculated by approximating the 3N-dimensional Cartesian coordinate probability distribution as a multivariate normal distribution. After aligning the simulation and correcting for symmetries (for details, see [12]) we calculated the atomic fluctuation covariance matrix of all non-hydrogen atoms, C, using *carma* [22]. From the correlation matrix, we calculate the differential entropy, H, of the corresponding multivariate distribution:

$$H = \frac{1}{2}\ln|2\pi e C| \qquad (15)$$

In order to estimate the error in our calculations due to autocorrelation in the simulation trajectoriess, we performed moving block bootstrapping [23]. For block sizes of 1, 500, 1000, 5000, and 10000, we generated 50 new trajectories from the original trajectories. For each information measure, we then calculated the standard error for each block size across the 50 corresponding trajectories. For all mean values, the mean for block size of 1 was reported. For all standard errors, the maximum error across all block sizes is reported.

### 3.3 Entropy Expansion

The number of n-body information terms to be calculated for the N-body expansion of the configurational entropy, is $(2^N - 1)(N - 1)$ after accounting for recursion. Each n-body term requires the determinant of a [dn x dn] matrix (where d is the dimension of each distribution, assuming all distributions are the same dimension), which has $O(n^3)$ complexity. Thus the calculation of the expansion has approximately $O(d^3 N^4)$ time complexity. We note that many of these terms are not unique, and the complexity can be reduced by storing and looking up terms.

All exponentials were fit using the nls function in R. All information measures were scaled by their appropriate scaling coefficient.

## 4 Results

The application of the *total intercorrelation* and entropy decomposition measures to characterize rigid-body behavior was illustrated with the analysis of the results from for 1 μs MD simulations of an all-atoms homology model of the serotonin (5-HT) receptor 5-HT$_{2A}$R, in the apo and 5-HT-bound states inserted in an explicit atomistic membrane model solvated with water and 0.15 mM NaCl. As described elsewhere [24], only the last 500 ns of each 1 μs trajectory was analyzed here to investigate the approximately equilibrated systems.

The analysis focused on the secondary structure of IL2 of the 5HT$_{2A}$R, considered to include residues I181-F186. Residues I181-F186 were helical within the initial structures of both states, and traditional secondary structure calculation using *stride* [25] indicates that the interior helical turn, composed of residues H182-R185, is stable throughout our simulations, with the turn being entirely helical for 84.3% of the apo trajectory and 89.7% of the 5-HT-bound trajectory.

We then calculated generalized correlation coefficients (Eq. 11) using *N-body mutual information* (Eq. 2) and *N-body total intercorrelation* (Eq. 8) for IL2 in both simulations to quantify the rigid-body behavior of the helical segment and to assess if there were differences in rigid-body behavior between the two states that could not be observed by calculating the secondary structure alone. We found that the apo state displayed weak rigid-body dynamics ($r_{mutual} = 0.30$ and $r_{inter} = 0.60$), while the 5-HT bound state displayed stronger rigid-body dynamics ($r_{mutual} = 0.52$ and $r_{inter} = 0.89$). These results indicate that there are more rigid-body motions in the 5-HT bound simulation, although both states have a helical segment in the IL2.

Using the entropy decomposition framework to analyze the dynamics of IL2, one would expect a high RBF (Eq. 13) and CO (Eq. 14) if the helical segment truly behaves as a rigid body helix, and a moderate RBF and low CO if the backbone is behaving like a rigid body but the side chains are not (likely a more accurate expectation based on the previously calculated generalized correlation coefficients). Conversely, if IL2 were behaving as a completely disordered segment, which is not expected from its helical secondary structure, RBF and CO would be low. We find that while IL2 is helical when the receptor is unbound, the RBF and CO parameters calculated from both the *mutual information* and *total intercorrelation* are low (see Table 1), indicating that IL2 contains a very flexible helix. In addition, we find that the RBF increases in the 5-HT bound state of the 5-HT$_{2A}$R. Interestingly, the comparison of CO$_{mutual}$ to CO$_{inter}$ reveals different trends upon 5-HT binding, indicating that the choice of information measure can influence the intepretation of the system's dynamics. Thus, most of the high-order correlation are identified as rigid body using *total intercorrelation*, but not when using *mutual information*. These results indicate that there is a significant increase in the rigidity of the IL2 upon ligand binding although the helical secondary structure is retained and comparable in both states.

**Table 1.** Rigid-body parameters of the apo and 5-HT bound 5-HT$_{2A}$R. The standard error is displayed in parenthesis.

|  | Apo | | 5-HT | |
| --- | --- | --- | --- | --- |
|  | mutual | inter | mutual | inter |
| **r** | 0.30 (0.002) | 0.60 (0.003) | 0.52 (0.002) | 0.89 (0.002) |
| **RBF** | 0.16 (0.002) | 0.50 (0.002) | 0.39 (0.002) | 0.90 (0.002) |
| **CO** | 0.77 (0.001) | 1.91 (0.001) | 1.11 (0.004) | 0.67 (0.005) |

Notably, a greater overall rigidity is indicated for both systems when using *total intercorrelation* as opposed to *mutual information*, as seen in the N-body generalized correlation coefficient and *rigid-body fraction*. We expect this result to be general and apply to other systems as well.

## 5 Discussion

In this work, we have developed and demonstrated new theoretical and computational tools for studying the rigid-body properties of protein secondary structure elements. From a theoretical basis, our new information measure, *total intercorrelation*, better describes the amount of correlated behavior between multivariate distributions, and we demonstrate this improvement by detecting rigid-body behavior in IL2 for the serotonin 5-HT$_{2A}$ receptor, a member of the class A GPCRs. Furthermore, through *total intercorrelation* and the entropy decomposition, we show that the discrete secondary structure categories used conceptually in structural biology cannot reveal true differences in the rigid-body properties of protein elements that can be significant for functional mechanisms. We find for the helical IL2 segments of the two studied systems, which turn out to have very different rigid-body characteristics, that the combination of *total intercorrelation* and entropy decomposition yields a more informative quantitative assessment of protein secondary structure than was obtainable previously.

We note that our analysis leads to previously unattainable insights regarding GPCR activation and ligand-dependent determination of signaling pathway. Thus, we find that IL2 of 5-HT$_{2A}$R transitions from a flexible helix to a more rigid-body helix upon binding the endogenous agonist 5-HT. As previous crystallography data [1–4] and computational analysis [8, 9] have pointed to the helix properties of IL2 in relation to GPCR activation for different pathways, it is possible that ligands can determine their agonist bias by allosterically modulating the rigid-body properties of IL2 upon binding. If this is so, it opens new opportunities for the design and discovery of drugs that modulate IL2 dynamics and thus its function, possibly even by binding to sites other than the traditional orthosteric and allosteric sites.